\documentclass[10pt]{iopart}

\usepackage{graphicx}

\newcommand{\prt}{\partial}
\newcommand{\la}{\lambda}

\graphicspath{{Figures/}}

\begin{document}

\title{Simple waves in a two-component Bose-Einstein condensate}

\author{S K Ivanov$^{1,2}$ and A M Kamchatnov$^{1,2}$}
\address{$^{1}$ Institute of Spectroscopy, Russian Academy of Sciences, Troitsk, Moscow, 108840, Russia}
\address{$^{2}$ Moscow Institute of Physics and Technology, Institutsky lane 9, Dolgoprudny, Moscow region, 141700, Russia}
\eads{\mailto{ivanoff.iks@gmail}, \mailto{kamchatnov@gmail.com}}

\begin{abstract}
We consider dynamics of simple waves in a two-component Bose-Einstein condensates. The evolution of the
condensate is described by the Gross-Pitaevskii equations which can be reduced for simple wave solutions
to a system of ordinary differential equations which coincide with those derived by Ovsyannikov for the two-layer fluid dynamics.
We solve the Ovsyannikov system for two typical situations of large and small difference between inter-species
and intra-species nonlinear interaction constants.
Our analytic results are confirmed by numerical simulations.
\end{abstract}

\pacs{67.85.Fg, 47.35.Fg}

\vspace{2pc}
\noindent{\it Keywords}: multicomponent Bose-Einstein condensates, simple waves, Riemann problem, dispersive shock waves, solitons, solitary waves, wave breaking, Whitham modulation equations

%\submitto{\JPB}

%\maketitle

\ioptwocol

\section{Introduction}

The study of multi-component nonlinear waves is one
of the fascinating topics which has potential applications
to dynamics of Bose-Einstein condensates (BECs)
\cite{KevrekidisFC-2008} and of optical pulses in fibers \cite{KivsharAgrawal-2003}.
The dynamics of such a condensate is much more complicated
compared with the one-component case. In particular, two types of motions are
possible in two-component BECs---``density
wave'' with in-phase motion of the components and ``polarization waves'' with counter-phase
their motion. It has been noticed \cite{QuPS-2016}
that the polarization dynamics can be separated from the
density dynamics even for the case of large amplitude waves, if the difference between intra- and
inter-species interaction constants is small.
In this case, the structures arising during the evolution of two-component condensate were
studied in \cite{IvanovKCPHI-2017} under the assumption that the total density is preserved
what is a good approximation in this case. In this paper, we go beyond this approximation and
study the dynamics without imposing any restrictions on the nonlinear interaction
constants except that we assume that the components can mix in the same volume of a trap.
To study typical situations, we derive differential equations describing the evolution of
density and flow velocity of
the condensate components for the case of the so-called ``simple wave solutions'' when
all physical variables depend on the single parameter.
This class of solutions can be applied to the problem of the evolution of an initial
discontinuity for two characteristic cases.
It will be shown that when the parameters of the condensate are close to the miscibility
boundary, the total density remains practically constant with very good accuracy.
For the situation of weak interaction of the condensate component, we find that
compound simple waves can be formed that
consist of merged one-component and two-component rarefaction waves.
Approximate solutions are found and their accuracy is confirmed by numerical simulations.

\section{The model}

One-dimensional dynamics of a two-component Bose-Einstein condensate
without the external potential
is described with a high accuracy
by the system of
the Gross-Pitaevskii (GP) equations
which can be written in
non-dimensional form as
\begin{equation} \label{GPE}
    {\rm i}\frac{\partial\psi_i}{\partial t}=-\frac{1}{2}
    \frac{\partial^2\psi_i}{\partial x^2}+g_{ii}|\psi_i|^2\psi_i+g_{ij}|\psi_{j}|^2\psi_i,
\end{equation}
where $i,j=1,2~(i\neq j)$ label the corresponding
condensate components, $(\psi_1,\psi_2)$ are the wave
functions of the components which are normalized
to the number of particles,
\begin{equation*}
    \int|\psi_i|^2dx=N_i,
\end{equation*}
so that $|\psi_i|^2=\rho_i$ is the density of particles in the $i$-th component.
The gradient of the phase $\varphi_i$ of the wave function
$\psi_i=\sqrt{\rho_i}\exp({\rm i}\varphi_i)$ is equal to
the flow velocity $u_i$ of $i$-th component.
Parameters $g_{ii}$ are the constants of interaction
between atoms of component $i$, and $g_{ij}$ are
the constants of interaction between atoms of different
species. Usually $g_{12}=g_{21}$, what we will assume in what follows.

If the phase $\varphi_i$ is
a single-valued function of coordinates, what means physically that
there are no vortices in the condensate, the wave functions
of the two-component condensate can be represented as
\begin{equation} \label{WaveFunction}
    \psi_i=\sqrt{\rho_i(x,t)} \exp\left({\rm i}\int^{x} u_i(x',t)dx'-{\rm i}{\mu_i}t\right),
\end{equation}
where $\mu_i$ is the chemical potential of the $i$-th component (see \cite{PitaevskiiStringari-2003}).
Substitution of (\ref{WaveFunction}) into (\ref{GPE}) and separation of real
and imaginary parts followed by differentiation of one
of the equations with respect to $x$ cast the GP equations to the so-called ``hydrodynamic form'':
\begin{equation} \label{Hydro}
    \eqalign{
    {\partial_t\rho_i}+{\partial_x}(\rho_iu_i)=0,   \cr
    {\partial_t u_i}  +u_i{\partial_x u_i}+ g_{ii}{\partial_x\rho_i } +g_{ij}{\partial_x\rho_j} \cr
    \qquad+{\partial_x}\left(\frac{({\partial_x}\rho_i)^2}{8\rho_i^2}
    -\frac{{\partial_x^2}\rho_i}{4\rho_i}\right)=0.
    }
\end{equation}
The first equation (\ref{Hydro}) provides conservation of
the number of particles in the corresponding condensate
component. If we drop out the last dispersion term in
the second equation (\ref{Hydro}), then we get
the Euler two-fluid hydrodynamics equations,
\begin{equation} \label{HydroDisp}
    \eqalign{
    {\partial_t\rho_i}+{\partial_x}(\rho_iu_i)=0, \cr
    {\partial_t u_i}+u_i{\partial_x u_i}+ g_{ii}{\partial_x\rho_i } +g_{ij}{\partial_x\rho_j}=0.
    }
\end{equation}
This system describes dynamics of condensates at characteristic scales
much greater than the healing length equal to unity in our non-dimensional variables.

\section{Simple waves}

In simple wave solutions the variables $\rho_i$, $u_i$ are assumed to depend on
the space and time coordinated via single function and it is convenient
to choose the characteristic velocity $c=c(x,t)$ as such a function.
Obviously, $c$ is a local velocity of the mode under consideration and therefore
it satisfies the equation
\begin{equation}\label{c-eq}
   \partial_t c+c \, \partial_x c=0.
\end{equation}
It it easy to find that the characteristic equation for the system (\ref{HydroDisp})
can be written in the form
\begin{equation} \label{CharacEq}
    (1-v_1^2)(1-v_2^2)=\frac{g_{12}^2}{g_{11}g_{22}},
\end{equation}
where, to simplify the notation, we have introduced
the variables $v_1$ and $v_2$ according to
\begin{equation} \label{}
    u_1-c=v_1\sqrt{g_{11}\rho_1}, \qquad u_2-c=v_2\sqrt{g_{22}\rho_2}.
\end{equation}
In the case when $g_{12}^2<g_{11}g_{22}$, what means physically that the components of the condensate are
miscible (see \cite{AoChui-1998}),
the left part of the characteristic equation (\ref{CharacEq}) is less than unity.
As was mentioned in Introduction, we confine ourselves to this situation only.

Since the variables $\rho_i$ and $u_i$ depend on the characteristic velocity $c$ only,
the equations (\ref{HydroDisp}) reduce to the system of ordinary differential equations
\begin{equation} \label{}
    \eqalign{
     \rho_iu_i'+(u_i-c)\rho_i'=0, \cr
     (u_i-c)u_i'+g_{ii}\rho_i'+g_{ij}\rho_j'=0,
    }
\end{equation}
where the prime denotes the derivative with respect to $c$
($\rho_i'=d\rho_i/dc$ and $u_i'=du_i/dc$). In terms of variables $v_i$ this system reads
\begin{equation} \label{}
     \eqalign{
     \rho_i v_i'+\frac{3}{2}v_i\rho_i'+\sqrt{\frac{\rho_i}{g_{ii}}}=0, \cr
     v_i \rho_i v_i'+\left(1+\frac{v_i^2}{2}\right)\rho_i'+\frac{g_{ij}}{g_{ii}}\rho_j'+\sqrt{\frac{\rho_i}{g_{ii}}}v_i=0.
     }
\end{equation}
Solving it with respect to derivatives,
we arrive at the system
\begin{equation} \label{BO}
    \eqalign{
    \frac{d\rho_1}{dc}=-\frac{2}{3}\sqrt{g_{22}(1-v_2^2)}\,f, \cr
     \frac{dv_1}{dc}=\frac{v_1}{\rho_1}\sqrt{g_{22}(1-v_2^2)}\,f-\frac{1}{\sqrt{g_{11}\rho_1}}, \cr
    \frac{d\rho_2}{dc}=\frac{2}{3}\sqrt{g_{11}(1-v_1^2)}\,f, \cr
     \frac{dv_2}{dc}=-\frac{v_2}{\rho_2}\sqrt{g_{11}(1-v_1^2)}\,f-\frac{1}{\sqrt{g_{22}\rho_2}},
    }
\end{equation}
where
\small
\begin{equation} \label{f}
    \eqalign{
    f(\rho_1,\rho_2,v_1,v_2) \cr
    \;\;=\frac{\sqrt{g_{22}\rho_1}\rho_2v_1(1-v_2^2)+\sqrt{g_{11}\rho_2}\rho_1v_2(1-v_1^2)} {\sqrt{g_{11}}g_{22}\rho_2v_1^2(1-v_2^2)^{3/2}-\sqrt{g_{22}}g_{11}\rho_1v_2^2(1-v_1^2)^{3/2}}.
    }
\end{equation}
\normalsize
We call these equations as Ovsyannikov equations since similar equations were first obtained
by him in the theory of two-layer shallow water dynamics \cite{Ovsyannikov-1979}.
These equations should be solved with the initial conditions
\begin{equation}\label{init}
    \eqalign{
    \rho_1|_{t=0}=\rho_{10}, \qquad v_1|_{t=0}= v_{10}, \cr
    \rho_2|_{t=0}=\rho_{20}, \qquad v_2|_{t=0}= v_{20}.
    }
\end{equation}
Typically, $v_{10}$ and $v_{20}$ can be found numerically
by solving the algebraic system which consists of the characteristic equation (\ref{CharacEq}) and the conditions
$c|_{t=0}\equiv c_0=u_{10}-v_{10}\sqrt{g_{11}\rho_{10}}=u_{20}-v_{20}\sqrt{g_{22}\rho_{20}}$,
where we have defined initial velocities as $u_1|_{t=0}\equiv u_{10}$ and $u_2|_{t=0}\equiv u_{20}$.
In important particular case when the initial flow velocities are equal to zero ($u_{10}=u_{20}=0$),
the parameters $v_{10}$ and $v_{20}$ are given analytically by simple formulas
\begin{equation} \label{}
    \eqalign{
    v_{10}^2
     =\frac{g_{11}\rho_{10}+g_{22}\rho_{20}\pm\sqrt{\Delta}} {2g_{11}\rho_{10}}, \cr
    v_{20}^2
     =\frac{g_{11}\rho_{10}+g_{22}\rho_{20}\pm\sqrt{\Delta}} {2g_{22}\rho_{20}},
     }
\end{equation}
where $\Delta=(g_{11}\rho_{10}-g_{22}\rho_{20})^2+4g_{12}^2\rho_{10}\rho_{20}$.
%\begin{equation*}
%\Delta=(g_{11}\rho_{10}-g_{22}\rho_{20})^2+4g_{12}^2\rho_{10}\rho_{20}.
%\end{equation*}
Here the upper sign ($+$) corresponds to the in-phase motion of the components
associated mainly with the total density oscillations, and the lower sign ($-$) corresponds to out-of-phase
motion of the components associated mainly with relative motion of the components in the ``polarization'' mode.
The system of the Ovsyannikov equations (\ref{BO}) can be easily solved numerically.
Below we apply it to the problem of evolution of an initial discontinuity in the flow
data, that is to the so-called Riemann problem.

\section{Solution of the Riemann problem}

We apply here the above developed theory to description of wave structures
evolving from initial discontinuities.
Let the initial conditions have a step-like form
\begin{equation} \label{}
    \eqalign{
    \rho_1|_{t=0}=\cases{\rho_1^L,  &for $x<0$\\
                        \rho_1^R,    &for $x>0$\\}, \cr
    \rho_2|_{t=0}=\cases{\rho_2^L,  &for $x<0$\\
                        \rho_2^R,    &for $x>0$\\}.
    }
\end{equation}
We assume that the initial flow velocities are equal to zero ($u_{10}=u_{20}=0$),
i.e. the components are at rest at $t=0$, and that the density of the second component
at the left boundary is equal to zero, too ($\rho_2^L=0$).
This means that we have a `vacuum' of the component $\rho_2$
for $x<0$ at the initial moment of time.
The dependence on the spatial coordinate and the time in this problem is self-similar, $c=x/t$,
since our initial conditions contain no parameters with the dimensions of length.
For brevity we denote $g_{11}=g_{22}=g$ and $g_{12}=\widetilde{g}$.
We consider two typical situations:
the case when the interaction between the components differs little from the interaction
of particles which belong to the same component ($g-\widetilde{g}\ll g$), and when the
intra-components and inter-components interactions are very different ($\widetilde{g}\ll g$).

\begin{figure}[t] \centering
\includegraphics[width=8cm]{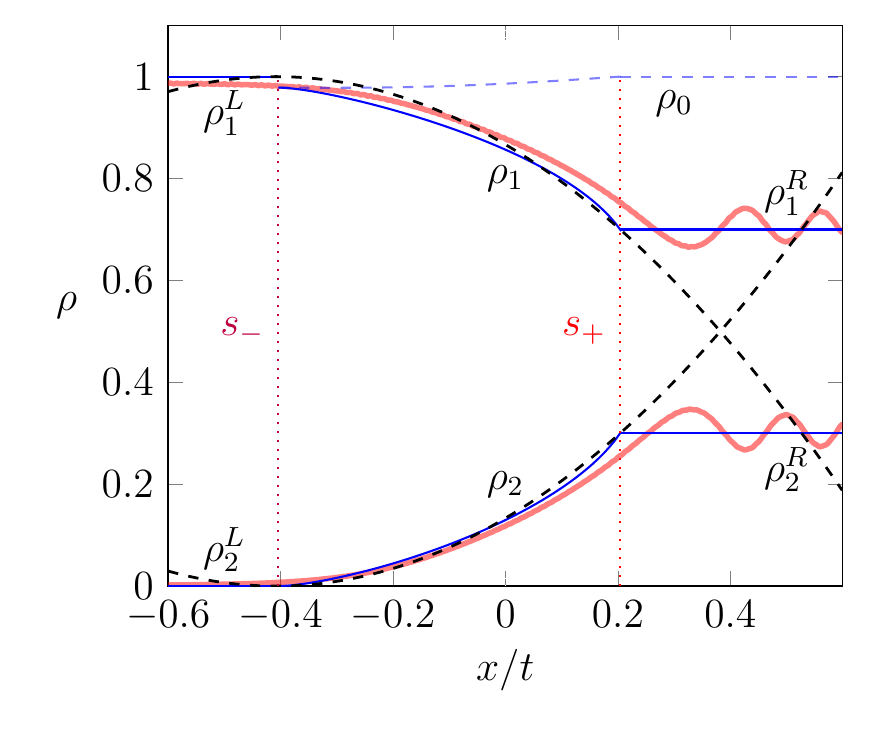}
\caption{In this plot we compare the numerical simulations for the density field $\rho(c=x/t)$
of the GP equations (\ref{GPE}) (red lines) with the numerical solution of the Ovsyannikov equations
(\ref{BO}) (blue lines) and the approximation (\ref{ApproxBo}) (black dashed lines).
The vertical dotted lines indicate velocities of the edges. The initial profile is characterized
by $\rho_1^L=1$, $\rho_2^L=0$, $\rho_1^R=0.7$, $\rho_2^R=0.3$ and the interaction constants are equal to
$g_{11}=g_{22}=g=1$, $g_{12}=\widetilde{g}=0.9$. The numerical solution of the Ovsyannikov equations
confirms the approximate constancy of the total density $\rho_0$ (blue dashed line) in the polarization mode.}
\label{Fig1}
\end{figure}

\subsection{Numerical solution of the Ovsyannikov equations}

First of all, we consider the case when $g-\widetilde{g}\ll g$.
This regime is of considerable practical interest.
For instance, it is realized in the condensate of $^{87}$Rb atoms in different states of the
hyperfine structure ($|1,-1\rangle$ and $|2,-2\rangle$) (see, e.g.,~\cite{VerhaarKK-2009}).
If $g-\widetilde{g}\ll g$, then the numerical solution yields
the distributions shown in figure~\ref{Fig1} by red lines.
Here in each component the rarefaction waves are formed.
These are simple waves in the polarization mode since they do not
practically affect the total density of the condensate.
In this flow one component replaces the another one leaving the total density practically constant,
$\rho_0=\rho_1^L+\rho_2^L=\rho_1^R+\rho_2^R$, as one can see
in the numerical solution of the Ovsyannikov equations (\ref{BO})
shown in the figure~\ref{Fig1} by a blue dashed line.
This property of the polarization mode in this regime was indicated in \cite{QuPS-2016,CongyKP-2016}
and our numerics serves as its additional justification.

Edges of the rarefaction wave propagate with velocities
\begin{equation} \label{}
    s_-=-\sqrt{\frac{8(g-\widetilde{g})\rho_{1}^R\rho_{2}^R}{\rho_{0}}}, \quad
    s_+=\sqrt{\frac{2(g-\widetilde{g})\rho_{1}^R\rho_{2}^R}{\rho_0}}.
\end{equation}
It is easy to see that in the limit $\widetilde{g}/g\rightarrow 1$
the velocities of the left and the right edges tend to zero.
In figure~\ref{Fig1} we illustrate such a structure and compare the numerical
solution of the GP equations (\ref{GPE}) with the numerical solution of the Ovsyannikov equations (\ref{BO}).
Black dashed lines correspond to the approximate analytical solution of the Ovsyannikov equations for the case
when the density of one of the components is much
greater than the density of the other component.
This analytical solution will be obtained below.

\begin{figure}[t] \centering
\includegraphics[width=8cm]{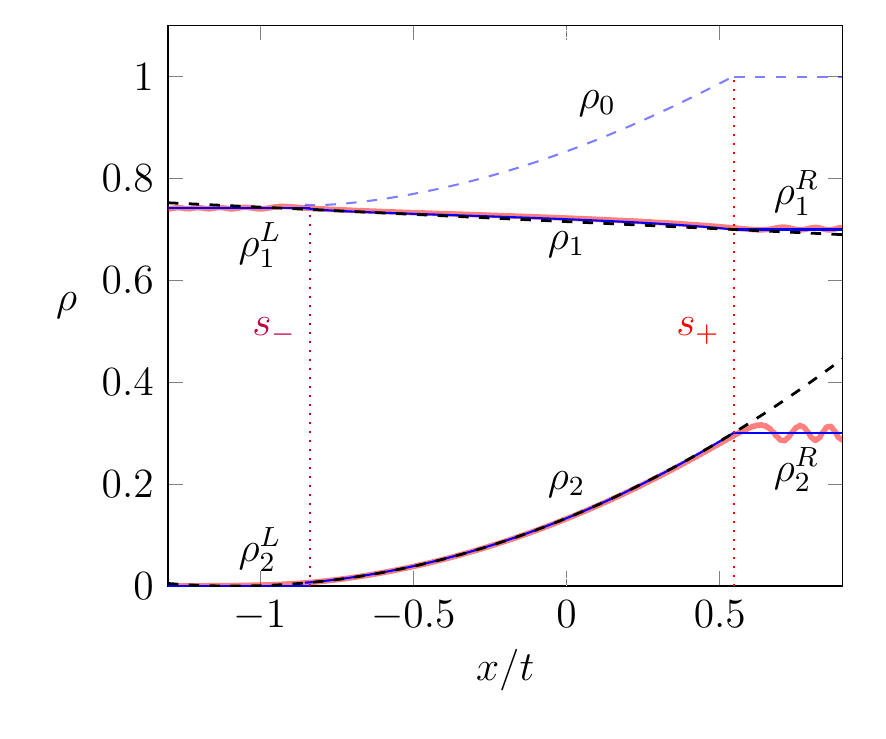}
\caption{We compare the density field $\rho(c=x/t)$ for the numerical solution of the GP equations
(\ref{GPE}) (red lines) with the numerical solution of the Ovsyannikov equations (\ref{BO}) (blue lines)
and the approximate analytical solution (\ref{ApproxBo}), (\ref{Ansatz}) (black dashed lines). The vertical dotted
lines indicate the edge velocities. The initial profile is characterized by $\rho_1^L=0.742$, $\rho_2^L=0$,
$\rho_1^R=0.7$, $\rho_2^R=0.3$ and the interaction constants are equal to $g_{11}=g_{22}=g=1$,
$g_{12}=\widetilde{g}=0.1$. The total density $\rho_0$ obtained by the numerical solution of the Ovsyannikov
equations is shown by a blue dashed line. It is clear that this density is inhomogeneous in this case.}
\label{Fig2}
\end{figure}

In the other case, when $\widetilde{g}\ll g$, we obtain distributions shown in figure~\ref{Fig2}.
Here interaction between the components is weak.
In this solution, the second component flows with formation of the rarefaction wave similar to
that for a one-component condensate flowing into vacuum.
The density of particles in this wave vanishes at the point $x=s_-t$,
where the velocity of the left edge is equal to $s_-=v_{20}\sqrt{g\rho_1^L}$.
Density of the first component at the left edge of the simple wave is
determined numerically from the continuity condition
$\rho_1^L=\rho_1(s_-)$. In the limit $\tilde{g}/g\rightarrow 0$,
the density of the first component remains constant along
a single rarefaction wave. One can say that in this case
the second component with a non-zero density at the right side of the
discontinuity flows into the first one, only slightly perturbing it in the flow region.
In figure~\ref{Fig2} we compare the numerical
solution of the GP equations (\ref{GPE}) with the numerical solution of the Ovsyannikov equations (\ref{BO})
for this type of the flow.
If the continuity condition at the left edge of the simple wave is violated,
then we get a formal multi-valued solution of the Ovsyannikov equations
what means that they lose here their applicability and we have to take into account dispersion
effects leading to formation of dispersive shock waves. This problem will be discussed below
(see figure~\ref{Fig3}).
The velocity of the right edge of the simple wave is equal to $s_+=-v_{20}\sqrt{g\rho_{2}^R}$.
It is worth mentioning that the edge velocities in the limit $\tilde{g}/g\rightarrow 0$ coincide
with the local sound velocities.

\subsection{Approximate analytical solution of the Ovsyannikov equations}

Now we suppose that one of the condensate components has a density
much less than the density of the other component.
To be definite, we assume that the first component has a greater density
than the second component ($\rho_1\gg\rho_2$).
In this case we can approximate $f$ given by equation (\ref{f}) as
\begin{equation} \label{}
    f\simeq-\frac{\sqrt{\rho_2}}{\sqrt{g_{11}g_{22}(1-v_1^2)}\,v_2},\qquad g_{22}\rho_2\ll {g_{11}}\rho_1.
\end{equation}
Substituting this expression in the second and fourth Ovsyannikov equations (\ref{BO}),
we find that $v_2=v_{20}=\mathrm{const}$, where $v_{20}$ is
defined by the initial condition. The density and the flow velocity of the second component are
equal respectively to
\begin{equation} \label{ApproxBo}
    \eqalign{
    \rho_2=\frac{1}{9g_{22}v_{20}^2} \left(2v_{20}\sqrt{g_{22}\rho_{20}}+u_{20}-c \right)^2, \cr
    u_2=\frac{1}{3} \left(2v_{20}\sqrt{g_{22}\rho_{20}}+u_{20}+2c \right).
    }
\end{equation}
Thus, in this approximation, we transform the system of four equations to a single equation
that determines the dynamics of the first component,
\begin{equation} \label{ApproxBoFor1}
    \frac{3}{2}\frac{v_1}{\rho_1}\frac{d\rho_1}{dc}+\frac{dv_1}{dc}+\frac{1}{\sqrt{g_{11}\rho_1}}=0.
\end{equation}
This equation can be reduced to the integral equation for the density and velocity of the first component
\begin{equation*}
    \rho_1^{3/2}v_1=\exp\left( \int_{c_0}^c\frac{dc}{v_1\sqrt{g_{11}\rho_1}} \right),
\end{equation*}
where the constant $c_0$ is determined by the initial conditions.

Let us return to the case when $g-\widetilde{g}\ll g$.
For $\rho_1\gg\rho_2$ the density of the second component can be described approximately
by the expression (\ref{ApproxBo}) with $u_{20}=0$.
Since the total density is constant,
the density of the first component is equal to $\rho_1(c)=\rho_0-\rho_2(c)$.
This density distribution is shown in figure~\ref{Fig1} by black dashed lines.
It should be noted that the approximate solution describes well the structure even for not
very large difference between $\rho_1$ and $\rho_2$.

In the case $\widetilde{g}\ll g$ the numerical solution suggests that the simple wave
of the first component of the condensate has approximately a linear form,
\begin{equation} \label{Ansatz}
    \rho_1=a+bc.
\end{equation}
Substituting this ansatz into expression (\ref{ApproxBoFor1}), we obtain the differential equation
\begin{equation} \label{}
    v_1'+\frac{3b}{2(a+bc)}v_1+\frac{1}{\sqrt{g(a+bc)}}=0,
\end{equation}
where the prime, as earlier, denotes the derivative with respect to $c$.
Solution of this equation with the initial condition $v_1(c_0)=v_{10}$ is given by
\begin{equation} \label{}
    v_1=\left[\frac{a+bc_0}{a+bc}\right]^{3/2}\left(v_{10}+\frac{\sqrt{a+bc_0}}{2b\sqrt{g}}\right)+ \frac{\sqrt{a+bc}}{2b\sqrt{g}}.
\end{equation}
This expression determines the distribution of flow velocity along the simple wave.
The parameter $b$ of the self-similar solution is the density derivative
with respect to the characteristic velocity $c$ ($\rho_1'=b$).
Then we can write the equation for $b$ with the use of the Ovsyannikov equations as
\begin{equation} \label{Ansatz:b}
    b=\frac{2}{3v_{20}}\sqrt{\frac{\rho_2(\overline{c})(1-v_{20}^2)}{g(1-v_1^2(\overline{c}))}},
\end{equation}
where we have introduced a convenient variable $\overline{c}=(s_-+s_+)/2$. Indeed,
generally speaking, the plot of the rarefaction wave of the first component has a nonzero curvature.
Therefore it seems reasonable to take the midpoint of this curve as a referent
point for calculation of the mean slope of the straight line approximation.
The constant $a$ can be found from the matching condition at the right edge of
the rarefaction wave where it matches with the right plateau ($\rho_1(c_0)={\rho}_1^R$),
\begin{equation} \label{Ansatz:a}
    a={\rho}_1^R+bv_{10}\sqrt{g{\rho}_1^R}.
\end{equation}
Thus, after substitution of (\ref{Ansatz:a}) into (\ref{Ansatz:b}),
we can find the parameters of the self-similar solution (\ref{Ansatz})
by numerical solution of algebraic equation (\ref{Ansatz:b}).
figure~\ref{Fig2} illustrates such an approximate structure where its comparison with the numerical
solution of the GP equations (\ref{GPE}) as well as with the numerical solution of the Ovsyannikov equations (\ref{BO})
and the approximate one (\ref{ApproxBo}), (\ref{Ansatz}) are also given. It is clear that
our approximate theory agrees with numerics very well.

\begin{figure}[t] \centering
\includegraphics[width=8cm]{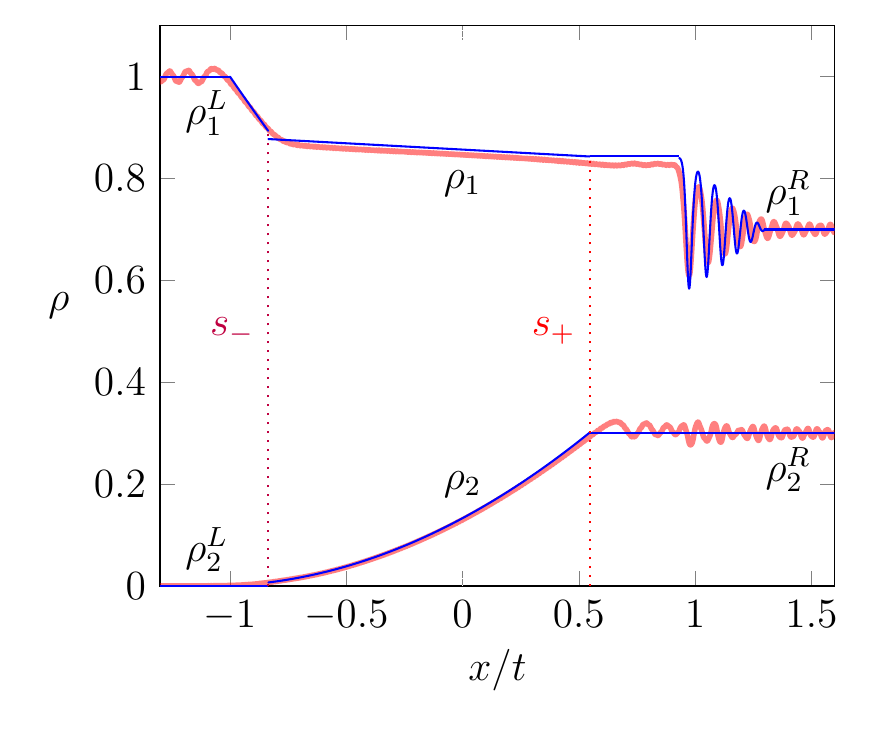}
\caption{In this plot we compare the numerical simulations for the density field $\rho(c=x/t)$
obtained by numerical solution of the GP equations (\ref{GPE}) (red lines) with the approximate solution (\ref{ApproxBo}),
(\ref{Ansatz}), (\ref{SingleRW}), (\ref{AproxPlataue}), (\ref{Periodic}) (blue lines).
The vertical dotted lines denote the velocities of the edges. The initial profile is characterized
by $\rho_1^L=1$, $\rho_2^L=0$, $\rho_1^R=0.7$, $\rho_2^R=0.3$ and the interaction constants are
equal to $g_{11}=g_{22}=g=1$, $g_{12}=\widetilde{g}=0.1$. The wave structure in the first component
consists of a dispersive shock wave on the right, two connected rarefaction waves on the left and
a plateau between them. The second component has a single rarefaction wave only.}
\label{Fig3}
\end{figure}

As was mentioned above, if $\widetilde{g}\ll g$ and the continuity condition
for the first component is not fulfilled ($\rho_1^L\neq\rho_1(s_-)$),
then the multi-valued region arises in the solution of the Ovsyannikov equations.
To consider such a situation, we assume here that the
initial parameters satisfy the inequality $\rho_1^L>\rho_1(s_-)$.
A typical density profile of the emerging wave structure is shown in figure~\ref{Fig3}.
According to this figure, in the second component there exists,
as before, a rarefaction wave only. However, the structure of the first
component becomes more complicated. This wave profile is similar to one
obtained in evolution of the initial discontinuity in a single component
condensate with boundary velocities equal to zero, where
a dispersive shock wave, that is the oscillatory wave structures emerging in evolution of
the condensate after wave breaking, is generated on the right side, a rarefaction wave on
the left side, and a plateau in between. However, there is one
difference between these two situations, namely, now the rarefaction wave is a composite one
because it consists of two regions: $(a)$ in the left one there is
no flow of the second component, and $(b)$ in the right region
where are rarefaction waves in both components.

First, we shall find a rarefaction wave in the region where the density
of the second component is equal to zero.
The solution of such a
problem is simplified considerably if we pass from the
ordinary physical variables $\rho_1$, $u_1$ for the first component to the so-called
Riemann invariants. For the equations (\ref{HydroDisp})
the Riemann invariants are well known and can be written as (see,
e.g.,~\cite{Kamchatnov-2000})
\begin{equation} \label{RiemInv}
    r_{\pm}=u_1\pm2\sqrt{g\rho_1}.
\end{equation}
The dispersionless system (\ref{HydroDisp}) can be written in the following diagonal
Riemann form,
\begin{equation} \label{DiagEq}
    \frac{\prt r_{\pm}}{\prt t}+V_{\pm}(r_{+},r_{-})\frac{\prt r_{\pm}}{\prt x}=0,
\end{equation}
where the ``Riemann velocities'' are given by
\begin{equation} \label{RiemVel}
    V_+=\frac34r_{+}+\frac14r_{-}, \qquad V_-=\frac14r_{+}+\frac34r_{-}.
\end{equation}
It is easy to express the physical variables in terms of $r_\pm$
\begin{equation} \label{}
    \rho_1=\frac{1}{16g}(r_{+}-r_{-})^2, \qquad u_1=\frac12(r_{+}+r_{-}).
\end{equation}
For the self-similar solution one has $r_{\pm}=r_{\pm}(c)$ and
the system (\ref{DiagEq}) reduces to
\begin{equation} \label{vp:vm}
    \frac{dr_\pm}{dc}\cdot\left(V_\pm-c\right)=0.
\end{equation}
A rarefaction wave is characterized by the fact that
one of the Riemann invariants has a constant value
along the flow. For the case shown in figure~\ref{Fig3}, the rarefaction
wave propagates to the left. Hence, the following invariant is constant
in it:
\begin{equation} \label{}
    r_{+}=u_1+2\sqrt{g\rho_1}=2\sqrt{g\rho_{1}^L},
\end{equation}
where we set its value equal to the value at the boundary
with the condensate at rest. The other Riemann invariant changes in such a way that the term
in parentheses in the equation (\ref{vp:vm}) with lower sign is equal to zero, $V_-=c$,
what yields
\begin{equation} \label{SingleRW}
    \eqalign{
    \rho_1=\frac1{9g}\left(2\sqrt{g\rho_{1}^L}-\frac{x}{t}\right)^2, \cr
    u_1=\frac23\left(\sqrt{g\rho_{1}^L}+\frac{x}{t}\right).
    }
\end{equation}
The left edge of this rarefaction wave propagates into the condensate
at rest with the local sound velocity.

Since the interaction between the condensate components is small, we assume
that the plateau and the dispersive shock wave
can be found approximately as the structures resulting from evolution of only one component,
i.e. we neglect here the interaction between the components.
Along the plateau the invariant $r_+$ must have
the same value as at the boundary between the rarefaction wave and the right plateau,
\begin{equation} \label{Plataue1}
    \overline{u}_1+2\sqrt{g\overline{\rho}_1}=2\sqrt{g\rho_{1}^L},
\end{equation}
where $\overline{\rho}_1$ and $\overline{u}_1$ are the density and flow velocity in the plateau.
After the passage through the dispersive shock
wave the Riemann invariant $r_-$ retains its value which gives the relation
\begin{equation} \label{Plataue2}
    \overline{u}_1-2\sqrt{g\overline{\rho}_1}=-2\sqrt{g\rho_{1}^R}.
\end{equation}
The equations (\ref{Plataue1}), (\ref{Plataue2}) allow one to find the densities and velocity of
the components in the plateau region:
\begin{equation} \label{AproxPlataue}
    \overline{\rho}_1=\frac14(\sqrt{\rho_{1}^L}+\sqrt{\rho_{1}^R})^2, \quad
    \overline{u}_1=\sqrt{g\rho_{1}^L}-\sqrt{g\rho_{1}^R}.
\end{equation}

Since the pioneering work of Gurevich and Pitaevskii (see \cite{GurevichPitaevskii-1973}), it
is known that wave breaking is
regularized by the replacement of the nonphysical multi-valued
dispersionless solution by a dispersive shock wave. This wave pattern
can be represented approximately as a modulated nonlinear periodic wave
\begin{equation} \label{Periodic}
    \eqalign{
    \rho_1=\frac{1}{4g}(\lambda_4-\lambda_3-\lambda_2+\lambda_1)^2 \cr
    \qquad+\frac1g(\lambda_4-\lambda_3)(\lambda_2-\lambda_1) \cr
    \qquad\times\mathrm{sn}^2(\sqrt{(\lambda_4-\lambda_2)(\lambda_3-\lambda_1)}\,\theta,m), \cr
    u_1=V-\frac{C}{g\rho_1},
    }
\end{equation}
where
\begin{equation}
    \eqalign{
    \theta=x-Vt,\quad V=\frac{1}{2}\sum_{i=1}^{4}\lambda_i, \cr
    m=\frac{(\lambda_2-\lambda_1)(\lambda_4-\lambda_3)}{(\lambda_4-\lambda_2)(\lambda_3-\lambda_1)},\quad 0\leq m\leq1; \cr
    C=\frac{1}{8}(-\lambda_1-\lambda_2+\lambda_3+\lambda_4) \cr
    \times(-\lambda_1+\lambda_2-\lambda_3+\lambda_4)
    (\lambda_1-\lambda_2-\lambda_3+\lambda_4);
    }
\end{equation}
and real parameters $\lambda_i$ are ordered according to the inequalities
\begin{equation*}
    \lambda_1\leq\lambda_2\leq\lambda_3\leq\lambda_4.
\end{equation*}
These parameters
are slow functions of $x$ and $t$ along a dispersive shock wave.
The periodic solution written in the form (\ref{Periodic}) has the
advantage that the parameters $\lambda_i$ are the Riemann invariants of the Whitham modulation equations,
and their evolution in our case is defined by the self-similar solution of the Whitham
equations presented in a diagonal Riemann form (see~\cite{ForestLee-1987,Pavlov-1987})
\begin{eqnarray} \label{Uisem}
    \frac{\partial\lambda_i}{\partial t}+V_i(\lambda_1,\lambda_2,\lambda_3,\lambda_4)\frac{\partial\lambda_i}{\partial x}=0,\quad i=1,2,3,4.
\end{eqnarray}
These equations describe the evolution of the parameters $\lambda_i$ and they can be derived by
averaging a proper number of conservation laws. This
method of deriving the modulation equations for
nonlinear waves was proposed by Whitham (see \cite{Whitham-1965,Whitham-1974}).
The velocities $V_i$ are expressed in terms of $K(m)$ and
$E(m)$, complete elliptic integrals of the first and second
kind, respectively (see \cite{ForestLee-1987,Pavlov-1987}). We need here only the expression for $V_3$,
\begin{equation} \label{}
    V_3=\frac{1}{2}\sum_{i=1}^{4}\lambda_i-
    \frac{(\lambda_4-\lambda_3)(\lambda_3-\lambda_2)K(m)}{(\lambda_3-\lambda_2)K(m)-(\lambda_4-\lambda_2)E(m)},
\end{equation}
As conserns the other Whitham velocities, it is important to notice that
in the soliton limit $m \to 1$ (i.e., $\la_3 \to \la_2$) the they  reduce to
\begin{equation} \label{SolLimit}
    \eqalign{
    V_1(\lambda_1,\lambda_2,\lambda_2,\lambda_4)=\frac32\lambda_1+\frac12\lambda_4, \cr
    V_4(\lambda_1,\lambda_2,\lambda_2,\lambda_4)=\frac32\lambda_4+\frac12\lambda_1,
    }
\end{equation}
In a similar way, in the small amplitude limit $m \to 0$ (i.e.,
$\la_3 \to \la_4$) we obtain
\begin{equation} \label{SmallLimit}
    \eqalign{
    V_1(\lambda_1,\lambda_2,\lambda_4,\lambda_4)=\frac32\lambda_1+\frac12\lambda_2, \cr
    V_2(\lambda_1,\lambda_2,\lambda_4,\lambda_4)=\frac32\lambda_2+\frac12\lambda_1.
    }
\end{equation}
This means that the edges of the dispersive shock wave
match the smooth solutions of the hydrodynamic dispersionless approximation.

From a formal point of view, we look again for the self-similar solutions
for the Whitham equations (\ref{Uisem}). Assuming that
the $\lambda_i$'s depend only on the variable $c=x/t$, we obtain at once
\begin{equation} \label{UisemSelfSim}
    \frac{d\la_i}{dc}\cdot\left(V_i(\la)-c\right)=0, \quad i=1,2,3,4.
\end{equation}
Hence we find again that only one Riemann invariant
varies along the dispersive shock wave, while the other three are constant.

From the matching conditions at the
edges of the dispersive shock wave we find that at the
soliton edge
\begin{equation} \label{}
    \lambda_1=\overline{r}_-/2, \quad \lambda_4=\overline{r}_+/2 \qquad \mbox{at} \qquad \lambda_3=\lambda_2,
\end{equation}
where $\overline{r}_{\pm}$ are the Riemann invariants of the dispersionless
theory that are defined by the equations (\ref{RiemInv}). Their values coincide with the
values in the plateau at the soliton edge of the
dispersive shock wave. Similarly, at the small-amplitude
edge we find
\begin{equation} \label{}
    \lambda_1=r_-/2, \quad \lambda_2=r_+/2 \qquad \mbox{at} \qquad \lambda_3=\lambda_4.
\end{equation}
Thus, the constant Riemann invariants are equal to
\begin{equation} \label{Lambda}
    \lambda_1=-\sqrt{g\rho_{1}^R}, \quad \lambda_2=\sqrt{g\rho_{1}^R}, \quad \lambda_4=2\sqrt{g\rho_{1}^L}.
\end{equation}
The $c$-dependence of $\lambda_3$ is determined
by the condition of vanishing of the expression in brackets in equation (\ref{UisemSelfSim}),
\begin{equation} \label{Lambda3}
    V_3\left(-\sqrt{g\rho_{1}^R},\sqrt{g\rho_{1}^R},\lambda_3,\frac12\overline{u}_1+\sqrt{g\overline{\rho}_1}\right)=\frac{x}{t}.
\end{equation}
Substitution of the values
of $\la_i$ resulting from (\ref{Lambda}) and (\ref{Lambda3})
into the periodic solution (\ref{Periodic}) yields the oscillatory
dispersive shock wave structure for the physical variables $\rho$ and $u$
shown in figure~\ref{Fig3}.

The derived formulas also give the analytical expressions
for the velocities of the edges of the dispersive
shock wave. The soliton edge and the small-amplitude edge
move, respectively, with the velocities
\begin{equation} \label{}
    \eqalign{
     s_-^{(DSW)}=\frac{\sqrt{g}}2\left(\sqrt{\rho_{1}^L}+\sqrt{\rho_{1}^R}\right), \cr
     s_+^{(DSW)}=\frac{\sqrt{g}(2\rho_{1}^L-\rho_{2}^R)}{\sqrt{\rho_{1}^L}}.
    }
\end{equation}
These values also agree well with the results of our
numerical calculation.

The rarefaction wave located in the region where the
second component of the condensate forms also a rarefaction wave
will be sought, as earlier, under the assumption that it has a linear shape (\ref{Ansatz}).
Here the parameter $a$ can be found from the matching condition of this rarefaction wave
with the plateau
\begin{equation} \label{}
    a={\rho}_1^R-b(\overline{u}_1-v_{10}\sqrt{g\overline{\rho}_1}).
\end{equation}
From this condition and the expression (\ref{Ansatz:b}) we
can find numerically both parameters of the ansatz (\ref{Ansatz}).
As one can see,
there is a slight difference between analytical results
and numerical simulations for the simple wave and the dispersive shock wave.
These deviations are associated with the neglect of interaction
between the components of condensate.

\section{Conclusion}

In this paper, we have derived the differential equations
witch describe the dynamics of simple waves in a two-component
Bose-Einstein condensate without imposing restrictions on
the relative interaction of these components.
The theory is applied to the Riemann problem of
evolution of an initial discontinuity for two specific cases of
relatively strong and relatively weak repulsion between the components.
In the first situation the rarefaction wave is formed  only.
It is found that in this situation the total density remains approximately uniform.
In the other case, the appearance of more complex structures with formation of
composite rarefaction waves, plateau and dispersive shock waves is demonstrated.
We have found an approximate solution of this system
where the density of particles in one of the components is
mach greater than in the other one.
Typical structures are described and it is shown that the analytical solutions are in
good agreement with the numerical results.

\end{document}